\documentclass[preprint,amsmath,amssymb,aps]{revtex4-1}

\usepackage{graphicx}
\usepackage{multirow}
\usepackage{color}
\usepackage{dcolumn}
\usepackage{bm}

\begin{document}


\title{Disruption of current filaments and isotropization of magnetic field in counter-streaming plasmas}

\author{C. Ruyer$^{1,2}$}
\email{charles.ruyer@cea.fr}
\author{F. Fiuza$^{1}$}
\email{fiuza@slac.stanford.edu}

\affiliation{$^1$High Energy Density Science Division, SLAC National Accelerator Laboratory, Menlo Park, California 94025, USA}
\affiliation{$^2$CEA, DAM, DIF, F-91297 Arpajon, France}

\begin{abstract}
We study the stability of current filaments produced by the Weibel, or current filamentation, instability in weakly magnetized counter-streaming plasmas. It is shown that a resonance exists between the current-carrying ions and a longitudinal drift-kink mode that strongly deforms and eventually breaks the current filaments. Analytical estimates of the wavelength, growth rate and saturation level of the resonant mode are derived and validated by three-dimensional particle-in-cell simulations. Furthermore, self-consistent simulations of counter-streaming plasmas indicate that this drift-kink mode is dominant in the slow down of the flows and in the isotropization of the magnetic field, playing an important role in the formation of collisionless shocks.
\end{abstract}

\maketitle
Collisionless shocks are ubiquitous in astrophysical plasmas and are known to be efficient particle accelerators \cite[]{Blandford_1987, Ackermann_2013}. Non-thermal acceleration requires the generation of waves and magnetic turbulence in the vicinity of the shock, which can efficiently scatter particles between the upstream and downstream regions, leading to a first order Fermi-type process \cite[]{Blandford_1987,Drury_1983,Malkov_2001,PRL_Matsumoto_2017}. Collective plasma phenomena play an important role in the slow down of the flows and in determining the nature of the field structure at the shock \cite{Treumann_2008,Bykov_2011}. The dominant plasma processes depend on both the flow and ambient medium conditions, and are not yet fully understood.
Particle-in-cell (PIC) simulations have significantly increased our ability to study the kinetic processes associated with shock formation and particle acceleration \cite[]{Kato_2008,Spitkovsky_2008,Martins2009,Bret_2013,POP_Bret_2014,PRL_Spitkovsky_2015}.
In particular, for initially weakly magnetized environments, it has been shown that the interpenetration of counter-streaming plasmas is mediated by the Weibel, or current-filamentation, instability (hereafter referred simply as Weibel instability) \cite[]{Weibel_1959,Fried_59,Davidson_1972}. 

The Weibel instability is triggered in an anisotropic momentum distribution configuration. In counter-streaming plasmas, it generates strong current filaments and amplifies magnetic field fluctuations of wavevectors transverse to the flow direction. It is known to be the fastest electromagnetic process for comparable flow densities and relativistic counter-streaming velocities, either for electron-ion or pair plasma  \cite[]{Bret_Gremillet_2010}. Recent experiments \cite{Fox_2013,Huntington_2013}, and earlier simulations \cite{Kato_2008,Fiuza_2012}, have also demonstrated that this instability can be dominant in the non-relativistic regime. While the linear phase of the instability has been widely studied, the nonlinear phase is not yet completely understood. After saturation, the current filaments can be unstable to secondary instabilities such as filament merging \cite[]{Honda_2000}. Previous theoretical and numerical studies indicate that during the early nonlinear phase, the filaments coalesce and lead to the increase of the dominant transverse magnetic wavelength \cite[]{Medvedev_2005,Achterberg_2007, Achterberg_2007b,POP_Ruyer_2015,PRL_Ruyer_16}.
However, the strong slow down of the ions and the onset of the turbulent ---  stochastic and nearly isotropic --- magnetic fields observed in the vicinity of the shock front remains unexplained. In particular, it is not yet clear which microscopic process mediates the longitudinal evolution of the current filaments and how that affects the slow down of the flows and the magnetic field structure of the shock.

In this Letter, we show for the first time that the current filaments produced by the Weibel instability are unstable due to a resonance between the drifting (current-carrying) ions and a drift-kink mode. This resonance occurs for a longitudinal wavelength that is significantly larger than the radius of the filament and the ion skin depth $c/\omega_{pi}$ ($c$ is the speed of light and $\omega_{pi}$ is the ion plasma frequency), and leads to the violent breaking of the filaments and slow down of the ions. Our analytical estimates for the resonant wavelength, growth rate, and saturation level of this instability are verified by 3D PIC simulations, both for a single filament case and for a large number of filaments generated self-consistently in the interaction of large-scale counter-streaming plasmas. Our simulations further show that the drift-kink mode is dominant in the isotropization of the magnetic field in these systems and must be taken into account for accurately describing shock formation and its structure.

In order to study the stability of current filaments produced in counter-streaming plasmas, we start by considering the current and magnetic field configuration at the saturation of the ion Weibel instability. At this stage, the current is primarily carried by the cold ions. The hot electron background provides a screening current and supports the magnetic pressure, such that $J_0 = en_0 (Z v_i -  v_e) =e  Zn_0 \kappa v_i$ and $n_0 k_B T_e \approx B_0^2/(2 \mu_0)$, where $\kappa = c/(\omega_{pe} R_0)$ is the approximate screening factor \cite[]{Achterberg_2007b,POP_Ruyer_2015}. Here, $n_0$, $R_0$, and $B_0$, are the filament plasma density, radius, and magnetic filed, $v_e (v_i)$ and $-e(eZ)$ are the electron (ion) velocity and charge, $T_e$ is the electron temperature, and $k_B$ and $\mu_0$ are the Boltzmann constant and the vacuum permeability. The Weibel instability saturates when the bounce frequency of the ions inside the filaments, $\omega_B = (2 \pi Ze B_0 v_i/(\lambda_B m_i))^{1/2}$, becomes comparable to the growth rate of the instability $\Gamma_W \simeq(v_i/c) \omega_{pi}$ \cite{Davidson_1972}, where $\lambda_B \equiv 4 R_0$ is the transverse wavelength of the magnetic field. Given that the most unstable wavelength is $\lambda_B \simeq c/\omega_{pi}$, the magnetization at saturation is $\sigma = B_0^2/(\mu_0 n_i m_i v_i^2) \approx 0.025$. Thus, at this stage, while the electrons are well magnetized, the ions remain unmagnetized.

Kinetic theory is needed to evaluate the stability of the filaments with $\lambda_B \sim k^{-1}\sim\rho_i$ (where $\rho_i$ is the ion gyroradius), but complete analytical solutions are challenging.
We note that similar problems arise in the stability analysis of a current sheet in magnetic reconnection \cite[]{Zhu_1993,Zhu_1996, Lapenta_97, Daughton_1998, Daughton_1999, Daughton1999b}. It is known from magnetohydrodynamics and two-fluid theory that under such conditions, the current structure can be unstable to kink-type modes \cite[]{Daughton_1999,apj_Mil_Nakar_2006}, which propagate with a phase velocity $\omega/k \approx v_i$ \cite[]{Daughton_1999}. Furthermore, this can give rise to resonances between the cold drifting ions and this mode, as noticed in the kinetic study of the drift-kink instability in current sheets \cite[]{Daughton_1998}, but their impact in these systems has not been addressed.

We will consider the role of the drift-kink mode and associated resonances in the longitudinal stability of current filaments mediated by the Weibel instability. More specifically, we will be looking for a resonance of the type $\omega = k_z v_i \pm \delta \omega$, where $k_z$ is the wavenumber along the flow direction, and $|\delta \omega| \ll |k_z| v_i$. This resonance can be particularly violent when $\omega \approx \omega_B$. In that case, while bouncing inside the filament, part of the ion longitudinal momentum is transferred to the transverse direction, creating  locally a transverse current ($J_{\perp,i}$). The cold and unmagnetized ions ($\rho_{i} \gtrsim R_0$), will then interact via $J_\perp \times B$, leading to bunching and coherent bouncing motion consistent with the kink deformation.
The distance between ions, $\delta z$, evolves as $ m_i\ddot{\delta z} \simeq J_{\perp}\delta B/n_0$, where $\delta B \simeq \mu_0 J_\perp \delta z$ is the magnetic field induced by $J_\perp \simeq \kappa J_{\perp,i}$ (neglecting the displacement current). We then obtain 
$\ddot{\delta z} \simeq \delta z(\mu_0 J_\perp^2 /eZ n_0 m_i)$, from which we can estimate the growth rate of the ion modulations $\Gamma_{kink} \simeq \kappa v_{\perp,i}\omega_{pi}/c$, where $v_{\perp,i} \simeq \omega_B R_0$ is the typical ion transverse velocity during the bouncing motion. We thus expect the resonance between bouncing ions and a drift-kink mode, to lead to strong deformation of the current filaments, with a wavelength
\begin{equation}
\lambda_\mathrm{kink} \simeq 2\pi \frac{v_i}{\omega_B} \simeq 2 \pi \alpha \sqrt{\frac{R_0\omega_{pe}}{c}}  \frac{c}{\omega_{pi}}, \label{eq:l0} 
\end{equation}
which has a growth rate
\begin{equation}
\Gamma_\mathrm{kink} \simeq \frac{1}{\alpha}\frac{v_i}{ c} \sqrt{\frac{m_e Z}{m_i}\frac{c}{R_0\omega_{pe}}} \omega_{pi},\label{eq:gf} 
\end{equation}
where $\alpha = 2/3 (1)$ in cylindrical (slab) geometry. We note that this resonance wavelength of current filaments produced by the Weibel instability can be significantly larger than the most unstable wavelength of the drift-kink mode in a current sheet, $k_{\rm max} \sim 1/R_0$ \cite{Daughton1999b}.

In order to validate our model for the deformation of current filaments due to ion resonance, we have performed 3D PIC simulations with the code OSIRIS 3.0 \cite{fonseca_osiris}. The simulations follow the evolution of a single current filament in a plasma, corresponding to the conditions near the saturation of the Weibel instability, as described above. The initial stable  configuration follows the same ideas of the well-studied Harris-type equilibrium \cite{Harris_62} and is obtained in our case by applying conservation of canonical momentum to a quasineutral plasma with a current profile $J_z = e n_0 \kappa v_i = J_0\cos[\pi r/(2R_0)]$ for $r=\sqrt{x^2+y^2}<R_0$, and $J_z = 0$ elsewhere. Ions are cold and electron screening is taken into account. The magnetic pressure is supported by the electrons and the plasma density is uniform outside the current filament. The simulations resolve the plasma with a cell size of $0.125c/\omega_{pe}$, a time step of 0.07 $\omega_{pe}^{-1}$, and used 8 particles/cell/species and periodic boundary conditions. A third order interpolation scheme is used for improved numerical accuracy. The transverse and longitudinal sizes of the simulation domain are $L_\perp = 64c/\omega_{pe} $  and $L_z =1024c/\omega_{pe}$ for $m_i/(Z m_e) = 100$.

\begin{figure}[ht!]
\includegraphics[width=0.64\columnwidth]{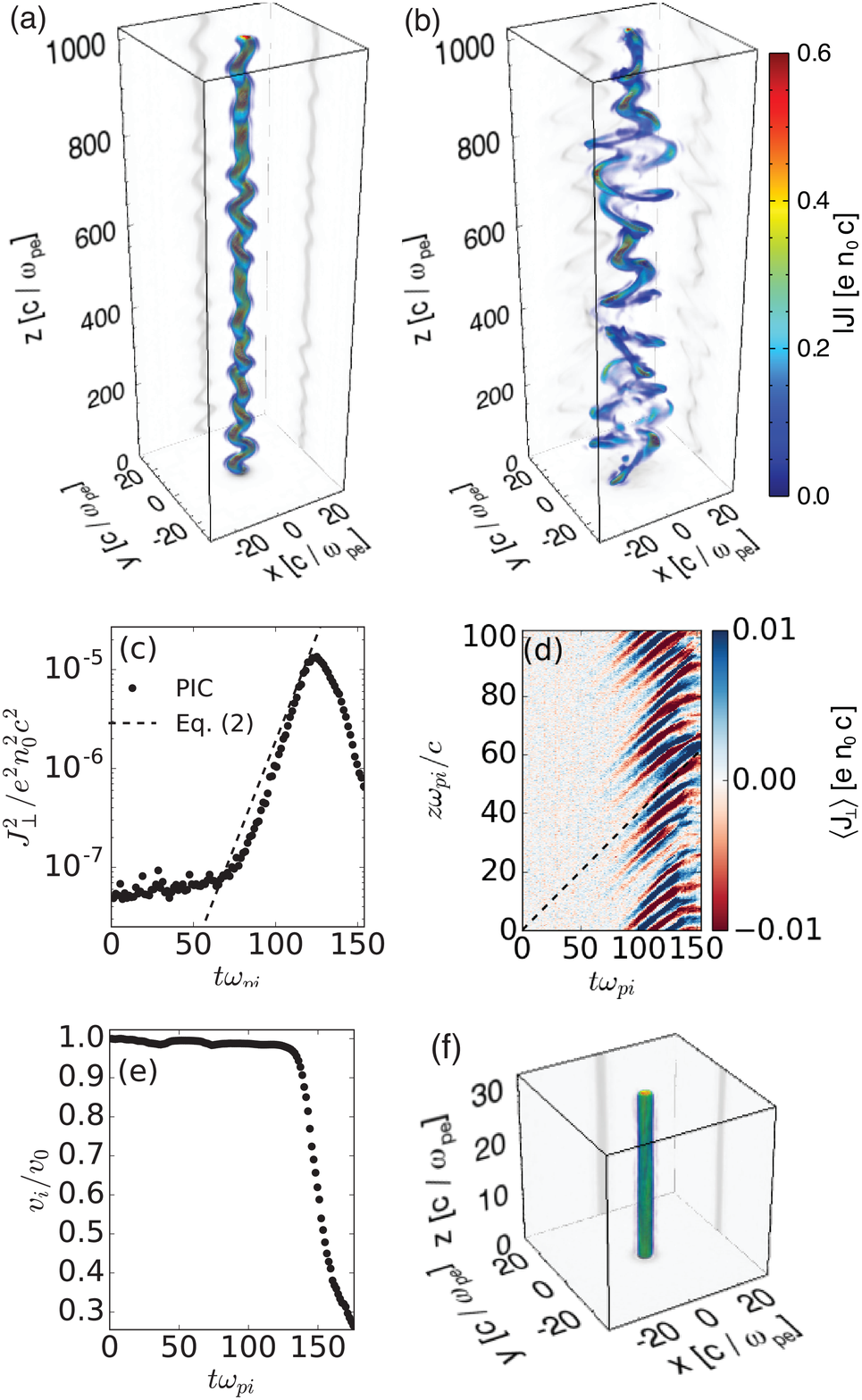}
\caption{\label{fig:kink} 
Evolution of the current density of a filament in a plasma with $m_i/m_e =100$, $R_0 = 3c/\omega_{pe}$, $J_0=0.5 e n_0 c$, and $L_z=  1024 c/\omega_{pe}$ at (a) $t = 114.2 \omega_{pi}^{-1}$ and (b), $t = 137.4 \omega_{pi}^{-1}$ (projections are shown in grey scale).
(c) Comparison of the growth rate of $J_\perp^2$ and Eq. \eqref{eq:gf} (dashed line). (d) Space-time diagram of the transverse component of the current, averaged over $r<R_0$. The dashed black line corresponds to $v = 0.4c$. 
(e) Temporal evolution of the maximum ion  drift-velocity. (f) Current density for the same parameters and color scale of (a,b) but with $L_z = 32 c/\omega_{pe}$, at $t = 314.9 \omega_{pi}^{-1}$.}
\end{figure}

Figure \ref{fig:kink}(a,b) illustrates the evolution the current density for a simulation with $R_0 = 3c/\omega_{pe}$ (corresponding to $\lambda_B = 1.2 c/\omega_{pi}$), $J_0 = 0.5 en_0 c$, and $T_{e,0}=0.4m_ec^2$ at the center of the filament. The growth of an $m = 1$ kink-type deformation on a time scale of $\sim 100\omega_{pi}^{-1}$ is clearly visible. The measured unstable wavelength is $7.8c/\omega_{pi}$, in good agreement with the prediction from Eq. \eqref{eq:l0}, $\lambda_{\rm kink} = 7.3 c/\omega_{pi}$. The growth rate is also well approximated by our estimate of Eq. \eqref{eq:gf} [Fig. \ref{fig:kink}(c)]. We further confirm that this kink perturbation has phase velocity comparable to the ion velocity, as shown in Fig. \ref{fig:kink}(d).  As the transverse displacement of the current becomes comparable to the initial filament radius, the distortions become strongly nonlinear, and the current is disrupted [Fig. \ref{fig:kink}(b)]. We observe that at this stage ($t \sim 130 \omega_{pi}^{-1}$), the growth of the kink modulation saturates and there is an abrupt decrease of the ion velocity (or current) [Fig. \ref{fig:kink}(f)]. This indicates that the kink-type deformation of the current filament is critical for the slow down of the flows.

In order to confirm the importance of the ion resonance, and not simply a fastest growing mode, we have repeated the simulation only changing the longitudinal box size to $L_z = 32 c/\omega_{pe} \simeq \lambda_{\rm kink}/2$. Figure \ref{fig:kink}(e) shows that in this case, the filament remains stable up to the maximum simulation time of $314.9 \omega_{pi}^{-1}$, indicating that indeed it is critical to capture the resonance wavelength.

\begin{figure}[h!]
\includegraphics[width=\columnwidth]{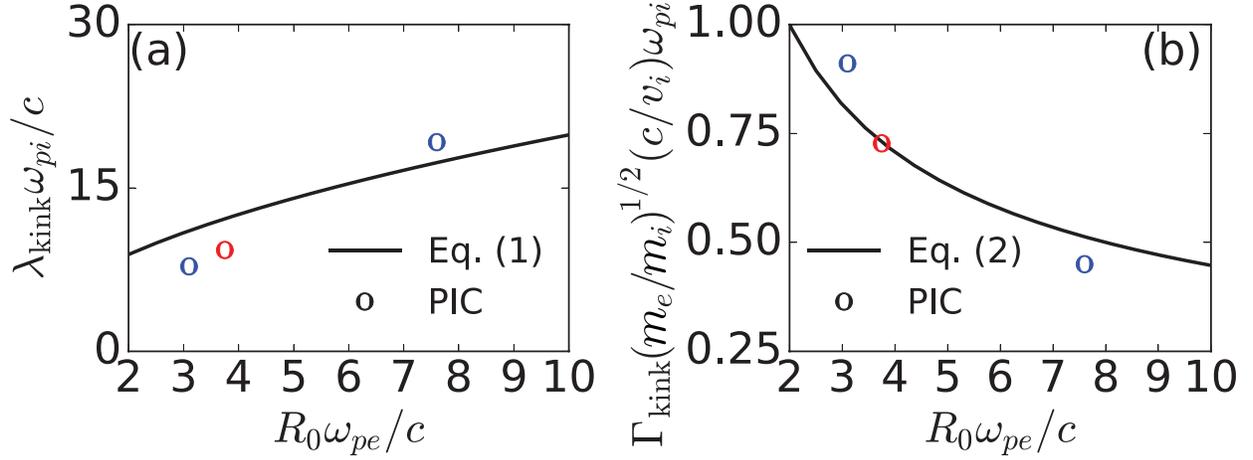}
\caption{\label{fig:linear}
Comparison of the (a) wavelength and (b) normalized growth rate ($\bar{\Gamma}_{\rm kink} = \Gamma_{\rm kink} ~\omega_{pi}^{-1} (m_i/m_e)^{1/2}(c/v_i)$) of the kink deformation of the current density obtained in 3D PIC simulations with the theoretical estimates of Eqs. \eqref{eq:l0} and \eqref{eq:gf}. Red and blue circles correspond to $m_i/m_e=25$ and $100$, respectively. The three simulations have initial filaments with radius and current density ($R_0 = 3c/\omega_{pe}, J_0 = 0.5e n_0 c$), ($R_0 = 3.5c/\omega_{pe}, J_0 = 0.5e n_0 c$), and ($R_0 = 7.5c/\omega_{pe}, J_0 = 0.08e n_0 c$).}
\end{figure}

Our estimates for the resonance wavelength and growth rate have been compared with additional 3D simulations for different $R_0$, $m_i/(Z m_e)$, and $v_i$. The results are illustrated in Fig. \ref{fig:linear}(a,b), showing good agreement with our model.

The kink-type deformations observed in a single filament should also be present and play an important role in the current-filamentation triggered self-consistently in counter-streaming plasmas. In order to confirm this, we have performed 3D simulations of the interpenetration of two symmetric, uniform, and cold electron-ion plasmas with $z$-aligned drift velocities. The counter-streaming flows are initialized with Maxwell-J\"uttner distributions with temperature $T_{e,i} = 1.28 \times 10^{-8}m_ec^2$, drift velocity $v_{e,i} = \pm 0.7c$, and mass ratio $m_i/m_e = 128$. We note that the high velocity and reduced mass ratio are chosen to optimize the growth time of the Weibel instability, given the large level of computational resources required by 3D simulations. We have used a time step $\Delta t=0.25/\omega_{pe}$ and a mesh size $\Delta x = \Delta y = \Delta z =0.5c/\omega_{pe}$. 

\begin{figure}[h!]
\includegraphics[width= \columnwidth]{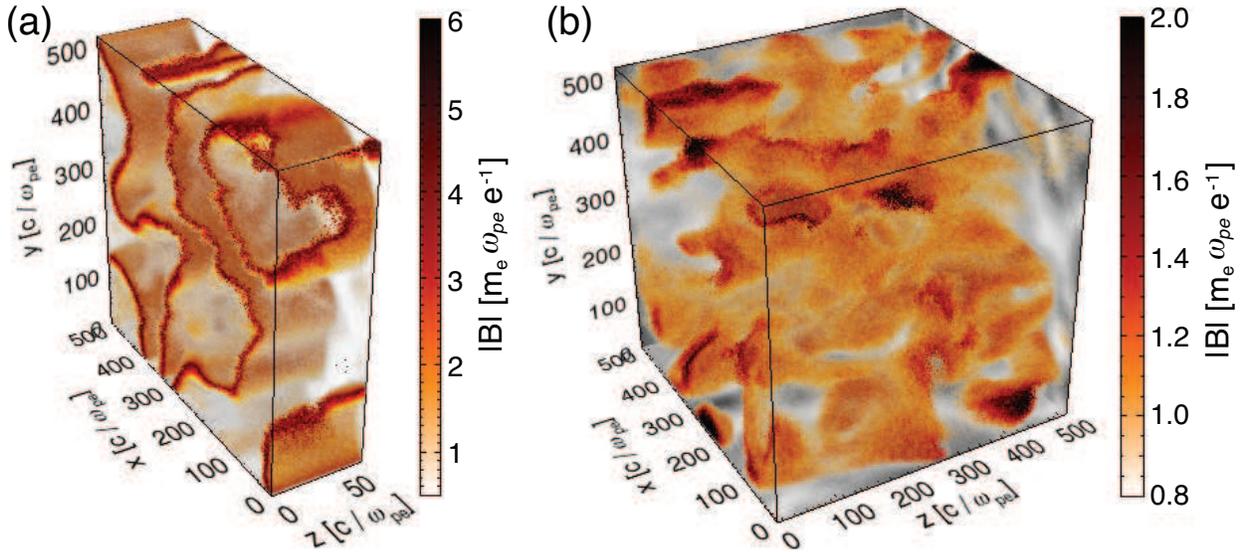}
\caption{\label{fig:periodic} 
3D simulations of the evolution of the magnetic field in the interaction of initially unmagnetized counter-streaming plasmas with $v_0 = 0.7c$ and $m_i =128m_e$, at $t = 400 \omega_{pi}^{-1}$. The longitudinal box size (along the flow direction) is (a) $L_z = 60c/\omega_{pe}<\lambda_\mathrm{kink}$ and (b) $L_z = 512c/\omega_{pe}>\lambda_\mathrm{kink}$.}
\end{figure}

Two 3D simulations have been performed in periodic geometry with $L_x = L_y =512c/\omega_{pe}$, one with  $L_z = 512c/\omega_{pe}$, the other with  $L_z = 60c/\omega_{pe}$. 
For both simulations, the electron-Weibel instability grows first, saturates and isotropizes the electron population in less than $10\omega_{pi}^{-1}$. 
After this period, the ion-Weibel instability dominates the system, saturating around $t\simeq 40\omega_{pi}^{-1}$. At this stage, the transverse wavelength of magnetic field is $\lambda_B \simeq 3 c/\omega_{pi}$. As discussed in Refs. \cite[]{Medvedev_2005, POP_Ruyer_2015,PRL_Ruyer_16}, for the simulation with a smaller longitudinal box, the nonlinear dynamics of the system is then dominated by the merging of filaments, which increases the dominant transverse magnetic wavelength while, heating the plasma. This is clearly observed in Fig. \ref{fig:periodic}(a) ($L_z = 60c/\omega_{pe}$), where filaments merge and their transverse wavelength approaches the system size by $t = 400 \omega_{pi}^{-1}$. No significant longitudinal perturbations of the filaments are observed. However, this changes drastically when the longitudinal simulation size is increased to $L_z = 512 c/\omega_{pe}$ [Fig. \ref{fig:periodic}(b)]. In this case, filament merging competes with strong longitudinal deformations associated with the drift-kink mode. After $t \simeq 200 \omega_{pi}^{-1}$, the coherent transverse magnetic modes associated with the Weibel instability are no longer observed. 

\begin{figure}[h!]
\includegraphics[width=\columnwidth]{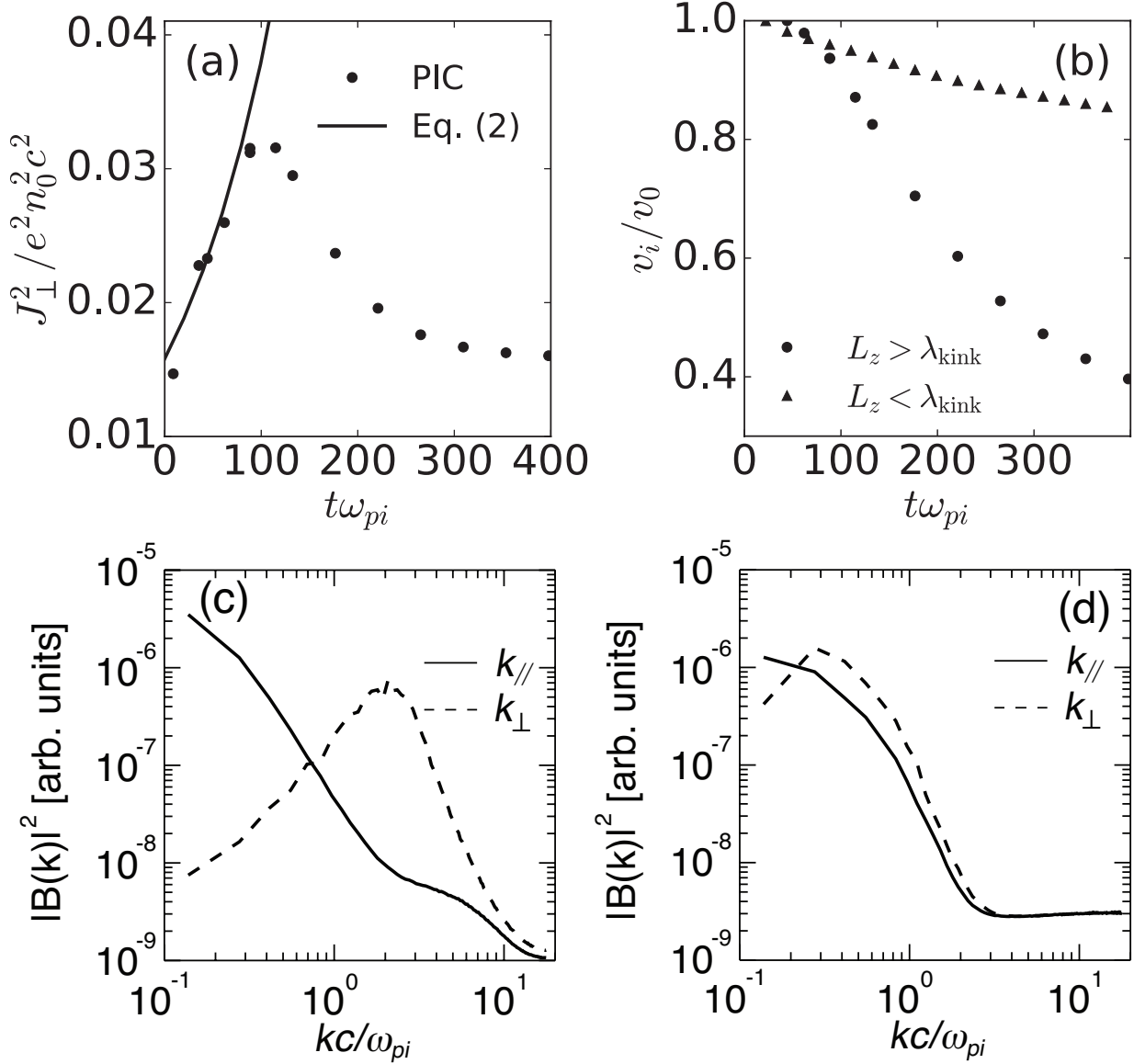} 
\caption{\label{fig:3dplasma} 
(a) Temporal evolution of the spatially averaged transverse current for the 3D counter-streaming plasma simulation of Fig. \ref{fig:periodic}(b) with $L_z = 512c/\omega_{pe}>\lambda_\mathrm{kink} $.
The corresponding  exponential growth $\propto \exp(2\Gamma_\mathrm{kink} t)$  is calculated with $\Gamma_\mathrm{kink}$ given by Eq. \eqref{eq:gf} and is plotted as the solid line.
(b) Temporal evolution of the  ion drift velocity for the simulations of Fig. \ref{fig:periodic}(a) (triangles) and Fig. \ref{fig:periodic}(b) (circles). Bottom plots show longitudinal (solid) and transverse (dashed) magnetic power spectrum (c) at saturation of the ion Weibel instability $t = 40 \omega_{pi}^{-1}$ and (d) after the drift-kink instability at $t = 220 \omega_{pi}^{-1}$ for the simulation of Fig. \ref{fig:periodic}(b).}
\end{figure}

We have analyzed the evolution of current density, and observe significant growth of its transverse component, consistent with kink modulations. The effective kink growth rate is in good agreement with our estimate from Eq. \eqref{eq:gf}, $\Gamma_\mathrm{kink}\sim 5.7\times 10^{-2} \omega_{pi}$ [Fig. \ref{fig:3dplasma}(a)]. The longitudinal spatial distribution of $J_\perp^2$, shows modulations with wavelength $\simeq 10 c/\omega_{pi}$. This is in very good agreement with the prediction of the resonance wavelength from Eq. \eqref{eq:l0}, $\lambda_\mathrm{kink} \simeq 12 c/\omega_{pi}$.

The kink-type deformations dominate the evolution of the system for $t > 100 \omega_{pi}^{-1}$ and are responsible for the rapid slow down of the flows [Fig. \ref{fig:3dplasma}(b)]. For $L_z < \lambda_{\rm kink}$, where the resonance is not captured, the slow down of the flows associated with filament merging occurs at a much slower rate. Furthermore, the longitudinal drift-kink mode also plays an important role in the isotropization of the magnetic field. This is illustrated by comparing the magnetic power spectrum at the saturation of the Weibel instability (before the growth of kink-type modulations) and after the saturation of the drift-kink mode [Figs. \ref{fig:3dplasma}(c,d)]. 

These results indicate that, in contrast with previous understanding, shock formation in weakly magnetized plasmas involves the combination of two instabilities. First, the Weibel instability produces strong current filaments and coherent near-equipartition magnetic fields. Then, the current filaments are unstable to the drift-kink mode, breaking, and causing the slow down of the flows and isotropization of the magnetic fields. The wavelength and growth rate of the kink-type deformation is well predicted by the resonance between the current-carrying ions and the propagating drift-kink mode. We note that even in scenarios where the resonance might not be present, \emph{e.g.} in pair-plasmas, the drift-kink instability is still expected to develop and play an important role in the nonlinear phase of the Weibel instability \cite[]{Vanthieghem}.

High-energy-density laser-plasma experimental platforms, which have been recently developed to study non-relativistic collisionless shocks \cite{Fox_2013,Huntington_2013,Ross_2017}, could probe the microphysics described in this work, in the near future. For the measured magnetic wavelength associated with the Weibel instability, $\lambda_B \simeq c/\omega_{pi}$ \cite{Huntington_2013}, the longitudinal kink-type modulations would develop with $\lambda_{\rm kink} \simeq 13.7 (A/Z)^{1/2} c/\omega_{pi}$, where $A$ is the atomic mass number of the ions. The expected time scale for the kink-type deformation of the current filaments is $\tau_{\rm kink} = \Gamma_{\rm kink}^{-1} \simeq 93.5 (A/Z)^{3/4}(c/v_i) \omega_{pi}^{-1}$. For typical flow velocity $v_i = 1000$ km/s, plasma density $n_0 = 10^{20}$ cm$^{-3}$, and $A/Z = 2$, this yields $\lambda_{\rm kink} \simeq 0.62$ mm and $\tau_{\rm kink} \simeq 5$ ns, which could be achieved at the OMEGA, NIF, or LMJ laser facilities.
 
In summary, we have found that the current filaments produced by the Weibel instability in counter-streaming plasmas are subject to violent kink-type deformations. This is associated with a resonance between the drift velocity of the current-carrying ions and a longitudinal drift-kink mode. Our analytical estimates for the resonance wavelength and growth rate of the kink perturbations were observed to be in good agreement with 3D PIC simulations. Furthermore, our simulations indicate that this drift-kink instability dominates the slow down of the flows and the isotropization of magnetic fields, thus mediating the formation of collisionless shocks. These findings can have important consequences for the generation of magnetic turbulence and for particle injection in weakly magnetized shocks.

\section*{Acknowledgments}
This work was supported by the U.S. Department of Energy SLAC Contract No. DE-AC02-76SF00515, by the U.S. DOE Office of Science, Fusion Energy Sciences under FWP 100237, and by the U.S. DOE Early Career Research Program under FWP 100331. The authors acknowledge the OSIRIS Consortium, consisting of UCLA and IST (Portugal) for the use of the OSIRIS 3.0 framework and the visXD framework. Simulations were run on Mira (ALCF) through INCITE and ALCC awards.

\bibliography{biblio}

\end{document}